\documentclass[12pt]{article}
\usepackage{graphicx,color,amsmath,amsxtra}
\usepackage{amssymb}
\usepackage[english]{babel}
\usepackage{amsfonts}

\newcommand\pubnumber{}
\newcommand\pubdate{\today}

\def\sendai{Department of Physics\\
Tohoku University, Sendai 980-8578, JAPAN}
\def\support{\footnote{Work supported by Grant-in-Aid for JSPS Fellows \#461001209. }}

\def\Title#1{\begin{center} {\Large #1 } \end{center}}
\def\Author#1{\begin{center}{ \sc #1} \end{center}}
\def\Address#1{\begin{center}{ \it #1} \end{center}}

\newcommand\pubblock{\rightline{\begin{tabular}{l} \pubnumber\\
         \pubdate  \end{tabular}}}
\newenvironment{Abstract}{\begin{quotation}  }{\end{quotation}}
\newenvironment{Presented}{\begin{quotation} \begin{center} 
             PRESENTED AT\end{center}\bigskip 
      \begin{center}\begin{large}}{\end{large}\end{center} \end{quotation}}
\def\Acknowledgements{\bigskip  \bigskip \begin{center} \begin{large}
             \bf ACKNOWLEDGEMENTS \end{large}\end{center}}

\def\beq{\begin{equation}}
\def\eeq#1{\label{#1}\end{equation}}
\def\eeqn{\end{equation}}

\def\beqa{\begin{eqnarray}}
\def\eeqa#1{\label{#1}\end{eqnarray}}
\def\eeqan{\end{eqnarray}}

\let\bar=\overbar

\def\Dslash{\not{\hbox{\kern-4pt $D$}}}
\def\dslash{\not{\hbox{\kern-2pt $\del$}}}

\def\msb{{\bar{\ssstyle M \kern -1pt S}}}

\begin{document}
\begin{titlepage}
\pubblock

\vfill
\Title{Cosmological perturbations in the models of dark energy and modified gravity}
\vfill
\Author{ Jiro Matsumoto\support}
\Address{\sendai}
\vfill
\begin{Abstract}
The quasi-static solutions of the matter density perturbation in various 
dark energy models and modified gravity models have been investigated in 
numerous papers. However, the oscillating solutions in those models have 
not been investigated enough so far. 
In this paper, the oscillating solutions, which have a possibility to unveil 
the difference between the models of the late-time accelerated expansion of the Universe, 
are also mentioned by using appropriate approximations. 

\end{Abstract}
\vfill
\begin{Presented}
CosPA 2013 \\
Symposium on Cosmology and Particle Astrophysics\\
12--15 November 2013, Honolulu, Hawai'i 96822
\end{Presented}
\vfill
\end{titlepage}
\def\thefootnote{\fnsymbol{footnote}}
\setcounter{footnote}{0}

\section{Introduction}
The fact that if the homogeneity of the Universe can be assumed then the Universe is accelerated expanding 
is clarified by the observations of Type Ia supernovae in 1990's\cite{SNIa}. 
We need to introduce some energy which have negative pressure to explain the accelerated expansion when we utilize the Friedmann equations which 
describe dynamics of the isotropic homogeneous universe. 
The energy introduced in this way are called dark energy. 
There are candidates of dark energy e.g. introducing the cosmological constant into 
the Friedmann equations, assuming the existence of the classical scalar field spreading over the whole universe, and so on. 
Whereas, there are modified gravity theories which can explain the accelerated expansion of the Universe not 
by introducing dark energy but by modifying the geometry of space-time or the gravitational constant. 
It is known that the $\Lambda$ Cold Dark Matter ($\Lambda$CDM) model, where $\Lambda$ means cosmological constant, 
is almost consistent with the observations of cosmic microwave background 
radiation, baryon acoustic oscillation, and type Ia supernovae. 
The $\Lambda$CDM model is regarded as the standard model of cosmology 
because it is simple besides is consistent with the observations. 
However, the other models of dark energy and modified gravity can realize an almost same expansion history of the Universe compared to that of the 
$\Lambda$CDM model. 
Therefore, we cannot determine the correct model which describes the real universe only from the growth history of the Universe. 
In this paper, it is mentioned whether or not differences between the models are always appeared 
by considering the evolution of the matter density perturbation 
as a perturbation from the background space-time of the Universe. 
The $\Lambda$CDM model, $k$-essence model\cite{k-essence 1,k-essence 2,k-essence 3} and 
$F(R)$ gravity model\cite{Buchdahl:1983zz,Nojiri:2006ri,Sotiriou:2008rp,DeFelice:2010aj,Nojiri:2010wj} will be considered 
as the typical models of dark energy and modified gravity. 

The cosmological perturbation theory is often used under the sub-horizon approximation, 
which consists of the two approximations in the small scale $a/k \ll 1/H$ 
and in the Hubble scale evolution $1/dt \sim H$, so that the perturbation should be consistent with the Newton gravity. 
However, the sub-horizon approximation is merely an approximation and is not always correct.   
In the following section, we will see what kinds of behaviors of the solutions are appeared when we 
do not adopt the quasi-static approximation $1/dt \sim H$. 
We use units of $k_\mathrm{B} = c = \hbar = 1$ and denote the gravitational constant $8 \pi G$ by
${\kappa}^2$ in the following. 

\section{Evolutions of the matter density perturbation in each model of dark energy and modified gravity}
The evolution equation of the matter density perturbation in the $\Lambda$CDM model is often expressed 
as follows: 
\begin{equation}
\ddot \delta + 2H \dot \delta - \frac{3}{2} \Omega_m H^2 \delta = 0,  
\label{e1}
\end{equation}
where $\delta \equiv \delta \rho / \rho$, $\rho$ is the energy density of the matter, 
$H$ is the Hubble rate defined by $\dot a (t) / a (t)$, and $\Omega_m$ is the matter fraction of the energy density of the Universe. 
Equation (\ref{e1}) is derived by using the sub-horizon approximation, whereas, 
if we do not use the sub-horizon approximation, then we obtain\cite{Matsumoto:2011ne} 
\begin{align}
\frac{d^2 \delta}{dN^2} + \Bigg [ 1&+ \frac{3}{2} (1+w) \Omega _m + 3(c_s^2 - w) \nonumber \\
& - \frac{d}{dN} \ln \bigg \vert - \frac{2k^4}{3a^4 H^2 \kappa ^2 \rho} 
+ 3(1+w) \bigg ( 1 + \frac{k^2}{3a^2 H^2} \bigg ) \bigg \vert \Bigg ]
\frac{d \delta}{dN} \nonumber \\
-& \Bigg \{ \frac{k^2}{3 a^2 H^2} (2+ 3w -3 c_s ^2 + 3w_\mathrm{eff}) 
+ 3(w-c_s ^2)- \frac{9}{2}(1+w)(w_\mathrm{eff}-w) \Omega_m \nonumber \\
 &- \bigg [ \frac{k^2}{3a^2 H^2} + 3(w - c_s ^2) - \frac{3}{2}(1+w) \Omega _m \bigg ] 
\frac{d}{dN} \ln \bigg \vert - \frac{2k^4}{3a^4 H^2 \kappa ^2 \rho} \nonumber \\
& \qquad \qquad \qquad \qquad \qquad \qquad \qquad \qquad 
+ 3(1+w) \bigg ( 1 + \frac{k^2}{3a^2 H^2} \bigg ) \bigg \vert \Bigg \}\delta = 0, 
\label{e2}
\end{align}
where $w$ is the equation of state parameter of the matter $w \equiv p/\rho$, $c_s$ is the sound speed $c_s ^2 \equiv \delta p/\delta \rho$, 
$k$ is the wave number, $a$ is a scale factor, and $N\equiv \ln a(t)$. $w_\mathrm{eff}$ is the effective equation of state parameter 
expressed as $w_\mathrm{eff} \equiv -2 \dot H / 3H^2 -1$. 
By expanding Eq.~(\ref{e2}) under the approximation $a/k \ll 1/H$ gives 
\begin{align}
\frac{d^2 \delta}{dN^2} &+ \left \{ \frac{1}{2} -6w +3c_\mathrm{s}^2 -\frac{3}{2} w_\mathrm{eff} 
+ O \left ( \Big ( \frac{k^2}{a^2 H^2} \Big ) ^{-1} \right ) \right \} \frac{d \delta}{dN} 
 + \left \{ \frac{c_\mathrm{s}^2 k^2}{a^2 H^2} \right.  \nonumber \\
+& \left. 3(w-c_\mathrm{s}^2)(1+3w +3w_\mathrm{eff}) - \frac{3}{2}(1+3w)(1+w_\mathrm{eff}) 
+ O \left ( \Big ( \frac{k^2}{a^2 H^2} \Big ) ^{-1} \right ) \right \} \delta  =0.
\label{e3}
\end{align}
It is found from Eq.~(\ref{e3}) that there are the wave number dependence of the matter density perturbation in the $\Lambda$CDM model, 
though it is sometimes said that the wave number dependence of the matter density perturbation is the peculiar property of 
$F(R)$ gravity model. 
As we have just seen, to evaluate the matter density perturbation without using the sub-horizon approximation can unveil 
some properties we have never known. 
In particular, the difference between the case the sub-horizon approximation is used and the case the sub-horizon approximation is not used 
is conspicuously appeared in $k$-essence model and $F(R)$ gravity model. 
In the following, we treat the equation of state parameter and the sound speed as $w=c_s=0$ 
by focusing on from the matter dominant era onwards. 

$k$-essence model is one of dark energy models, and its action is described by 
\begin{align}
S=  \int d^4 x \sqrt{-g} \bigg \{ \frac{R}{2\kappa^2} - K(\phi, X) 
+ L_\mathrm{matter} \bigg \} ,\quad X \equiv - \frac{1}{2}\partial^\mu \phi \partial_\mu \phi\, .
\label{e4}
\end{align}
Here $\phi$ is a scalar field and $L_\mathrm{matter}$ expresses the Lagrangian density of the matter. 
In $k$-essence model, the evolution equation of the matter density perturbation is not two dimensional but four dimensional 
because the number of the parameters in the Einstein equation are increased by the existence of the scalar field\cite{Bamba:2011ih}. 
We can decompose the four dimensional equation into the following two dimensional equation (\ref{e5}) and 
the solution (\ref{e6}) by considering that the scale of the density fluctuation we can observe is much less than 
the horizon scale of the Universe $a/k \ll 1/H$. The equation is given as 
\begin{align}
c_\phi ^2 \left \{ \frac{d^2 \delta}{dN^2} + \left( \frac{1}{2} - \frac{3}{2}w_\mathrm{eff} \right ) \frac{d \delta}{dN} 
-\frac{3}{2} \Omega_\mathrm{m} \delta \right \}= 0, 
\label{e5}
\end{align}
where $c_\phi$ is the sound speed in $k$-essence model defined by 
$c_\phi ^2 \equiv (p_{\phi})_{,X} / (\rho_{\phi})_{,X} = K_{,X}/(K_{,X}+ \dot \phi ^2 K_{,XX})$\cite{k-essence 2}. 
Here, $\rho_\phi$ and $p_\phi$ are the energy density and the pressure of the scalar field, respectively. 
The subscript $_{,X}$ means derivative with respect to $X$. 
The solution is expressed by 
\begin{align}
\delta _\mathrm{oci}(N) = & C(N) \cos \left [\int ^N dN' \frac{c_\phi k}{a H} \right]
+ r_1 C(N) \sin \left [\int ^N dN' \frac{c_\phi k}{a H} \right]\, , \label{e6} \\
\frac{d}{dN} \ln \vert C(N) \vert =& \frac{d}{dN} \ln \vert \dot \phi 
\vert - \frac{3}{4} \frac{d}{dN} \ln \vert K_{, X} \vert
\nonumber \\ 
&+ \frac{1}{4} \frac{d}{dN} \ln \left\vert K_{,X} + \dot \phi ^2 K_{, XX} \right\vert 
+ \frac{d}{dN} \ln \vert 4K_{, X} + \dot \phi ^2 K_{, XX} \vert\, , 
\label{e7} 
\end{align}
where $r_1$ is an arbitrary real constant. 
Equation (\ref{e5}) is equivalent to Eq.~(\ref{e3}) in the leading terms when $c_\phi$ is not vanished. 
On the other hand, the oscillating solution represented by Eq.~(\ref{e6}), which cannot be realized in the $\Lambda$CDM model, 
is peculiarity of $k$-essence model. 
The behavior of the oscillating solution depending on the form of the function $K(\phi,X)$ can be decaying or growing. 
Therefore, we should evaluate the behavior of the solution by calculating the effective growth factor represented by 
Eq.~(\ref{e7}) in each model. 

Next, we consider the following action as $F(R)$ gravity model, 
\begin{align}
S=\frac{1}{2 \kappa ^2} \int d^4 x \sqrt{-g} \left [ R+f(R) \right ]+S_\mathrm{matter}, 
\label{e8}
\end{align}
where $f$ is an arbitrary function of the scalar curvature $R$, and $f(R)$ 
represents the deviation from the Einstein gravity. 
It is known that $F(R)$ gravity model is equivalent to the scalar field model, which has a non-minimal coupling between 
the scalar field and the matter. 
Therefore, the evolution equation of the matter density perturbation is expected to be four dimensional same as in $k$-essence model. 
In fact, it is shown in \cite{delaCruzDombriz:2008cp} that the evolution equation is four dimensional in $F(R)$ gravity model. 
The coefficients of the equation are, however, too complicated to be definitely written down, 
so we need to expand the coefficients by applying the approximations $f_R \equiv df(R)/dR \ll 1$ and 
$a/k \ll 1/H$. 
Then, it is necessary to careful which approximations we should give priority to. 
In the following, $f_R, Rf_{RR}, RRf_{RRR} \ll 1$, where subscripts $_R$ means derivative with respect to $R$, 
take priority over $a/k \ll 1/H$ to describe the expansion history of the Universe 
similar to that of the $\Lambda$CDM model. 
The four dimensional equation is, then, expressed as follows\cite{Matsumoto:2013sba}: 
\begin{align}
\delta '''' &+ \left \{ \frac{ 12\mathcal{H}^2 (-2+ \mathcal{H}''/\mathcal{H}^3)f_{RRR}}{a^2f_{RR}}
+\frac{1-\mathcal{H}'/\mathcal{H}^2}{-2+ \mathcal{H}''/\mathcal{H}^3} + O(\mathcal{H}^2/\chi^2) \right \} \mathcal{H}
 \delta ''' \nonumber \\
&+ \chi^2 \bigg \{
\left ( 1 + O(\mathcal{H}^2/\chi^2) \right ) \delta '' 
+ \mathcal{H} \left ( 1 + O(\mathcal{H}^2/\chi^2) \right )  \delta ' \nonumber \\
& \qquad \qquad \qquad \qquad \qquad \qquad + \mathcal{H}^2  \left ( 2 \frac{\mathcal{H}'}{\mathcal{H}^2}-  \frac{\mathcal{H}''}{\mathcal{H}^3} 
 + O(\mathcal{H}^2/\chi^2) \right ) \delta  \bigg \} =0 
\label{e9}, \\
& \chi \equiv \sqrt{\frac{a^2}{3f_{RR}}\frac{1-\mathcal{H}'/\mathcal{H}^2}{2- \mathcal{H}''/\mathcal{H}^3}}.
\label{e10}
\end{align}
where $\mathcal{H} \equiv a'/a$ and the prime means the derivative with respect to the conformal time $\eta = \int dt/a$. 
Noting to the terms proportional to $\chi^2$, we obtain 
\begin{equation}
\frac{d^2 \delta}{dN^2} + \left ( \frac{1}{2} - \frac{3}{2}w_\mathrm{eff} \right ) \frac{d \delta}{dN}
 +\left ( 2\frac{\dot H}{H^2}+\frac{\ddot H}{H^3} \right ) \delta =0. 
\label{e11}
\end{equation}
Equation (\ref{e11}) is equivalent to Eq.~(\ref{e3}) 
when the $R$ derivatives of $f(R)$ are little. 
On the other hand, if we use the WKB approximation under the condition $\vert \chi \vert \gg 1$ then we have 
\begin{equation}
\delta (\eta) = C_1 {\rm e}^{\int f_\mathrm{eff} dN +i \int \chi d \eta} 
+ C_2 {\rm e}^{\int f_\mathrm{eff} dN -i \int \chi d \eta}, 
\label{e12}
\end{equation}
where $C_1$ and $C_2$ are arbitrary constants, and the effective growth factor $f_\mathrm{eff}$ 
is defined as 
\begin{equation}
f_\mathrm{eff}=1-\frac{5}{2} \frac{d}{dN} \ln \vert \chi \vert -2 \frac{d}{dN} \ln \vert f_{RR} \vert
 + \frac{1-\mathcal{H}'/\mathcal{H}^2}{2-\mathcal{H}''/\mathcal{H}^3}. 
\label{e13}
\end{equation}
Considering the Friedmann equation and the condition $f_R, Rf_{RR}, RRf_{RRR} \ll 1$, we can simplify Eq.~(\ref{e13}) into 
\begin{align}
f_\mathrm{eff} \simeq -\frac{1}{2} + \frac{9}{2} \left ( 2-\frac{\mathcal{H}''}{\mathcal{H}^3} \right ) \frac{\mathcal{H}^2 f_{RRR}}{a^2f_{RR}}. 
\label{e14}
\end{align}
Here, $\mathcal{H}''/\mathcal{H}^3\simeq 1/2$ is held in the matter dominant era. 
Whereas, viable models of $F(R)$ gravity are generally satisfies the condition $f_{RR}>0$ imposed from the quantum stability. 
Therefore, the behavior of the oscillating solution is determined by the sign of $f_{RRR}$. 
If the form of $f_{RR}$ is described by negative power law of $R$ or exp$(-\alpha R)$, $\alpha >0$, then 
$f_{RRR}<0$ and $f_\mathrm{eff}$ becomes negative. 
That is to say, the behavior of the matter density perturbation is determined by the quasi-static solution 
because the other solution is decaying oscillating solution. 
In this case, it is difficult to find the difference between $F(R)$ gravity model and the $\Lambda$CDM model from 
the matter density perturbation. 
In fact, famous viable models of $F(R)$ gravity have such a behavior, 
so we can make models which cannot be distinguished from the $\Lambda$CDM model by the observations concerned with 
the background and the perturbative evolution of the Universe. 

\section{Summary}
The following behaviors of the matter density perturbation in the models of dark energy 
and modified gravity are unveiled by considering them without applying the subhorizon approximation. 
In the $\Lambda$CDM model, the wave number dependence of the matter density perturbation is 
appeared in sub-leading terms. 
There is not only the quasi-static solution but also the oscillating solution which 
can give unignorable contributions in $k$-essence model. 
Although there is the oscillating solution in $F(R)$ gravity, 
viable F(R) gravity models cannot be distinguished from the $\Lambda$CDM
model by evaluating the growth rate of the structure formation when
we fit their background evolution to the observational results.
A sufficient conditions for the fast fluctuating mode to be the decaying
oscillating solution are $f_{RR} > 0$ and $f_{RRR} < 0$.
\Acknowledgements
The author is grateful to Shin'ichi Nojiri for advices to 
this investigation.

\end{document}